\documentclass{aa}
\usepackage{graphics}
\usepackage{psfig}
\usepackage{placeins}

\def\etal{{et\,al.~}}

\def\erga{\hbox{erg cm$^{-2}$ s$^{-1}$ \AA$^{-1}$}}

\def\msun{M\ifmmode _{\odot}\else$_{\odot}$\fi}
\def\mwd{M\ifmmode _{\rm wd}\else$_{\rm wd}$\fi}
\def\msec{M\ifmmode _{\rm sec}\else$_{\rm sec}$\fi}
\def\rsun{R$_{\odot}$}

\def\degs{\ifmmode ^{\circ}\else$^{\circ}$\fi}
\def\amin{\ifmmode ^{\prime}\else$^{\prime}$\fi}
\def\asec{\ifmmode ^{\prime\prime}\else$^{\prime\prime}$\fi}

\def\v13{V1309~Ori}
\def\rxj05{RX J0515.41+0104.6}

\def\kmps{km\,s$^{-1}$}
\def\heII{He\,{\sc ii}\,$\lambda$4686}
\begin{document}

\title{System parameters of the long-period polar V1309~Ori\thanks{Based partially on observations with the NASA/ESA {\it{Hubble Space Telescope}} obtained at the Space Telescope Science Institute, which is operated by the Association of Universities for Research in Astronomy, Inc., under contract NAS 5-26555}}

  \author {
	A.~Staude
	\and
	A.~D.~Schwope\thanks{Visiting 
                astronomer,
              German-Spanish Astro\-no\-mi\-cal Center, Calar Alto, 
              operated by the Max-Planck-Institut f\"ur Astronomie, 
              Heidelberg, jointly with the Spanish National Commission 
              for Astronomy.}
	\and
	R.~Schwarz{$^{\star\star}$}
	}
   \offprints{A.~Staude ({\tt astaude@aip.de})}
                               
  \institute {
	Astrophysikalisches Institut Potsdam, An der Sternwarte 16,
	D--14482 Potsdam, Germany
}
\date{Received Feb. 16, 2001; accepted May 15, 2001}

\abstract{
Based on high-resolution optical spectroscopy in the blue and the 
near infra-red spectral range, we derived velocity images (Doppler 
tomograms) of the mass-donating secondary star and the accretion 
stream of the long-period eclipsing polar \v13 (\rxj05).
Combined with HST-spectroscopy of high time resolution and optical 
photometry we were able to derive the main system parameters
and to determine the accretion geometry of the binary. The 
length of the eclipse of the white dwarf is $\Delta t_{ecl} = 2418\pm60$\,s,
the mass ratio $Q = 1.37 - 1.63$, and the orbital inclination $i=76.6\degr -
78.9\degr$. The surface of the secondary star could be resolved in 
the Doppler image of NaI absorption lines, where it shows a marked
depletion on the X-ray irradiated side. The accretion 
geometry proposed by us with a nearly aligned rotator, co-latitude 
$\delta \simeq 10\degr$,
tilted away from the ballistic stream, azimuth $\chi \simeq -35\degr$,
explains the shape of the emission-line Doppler tomograms and the shape 
of optical/UV eclipse light curves.   
}

\maketitle

\section{Introduction}

\begin{table*}[htb]
\small
\caption{The data sets of V1309 Ori used in this paper. (MLO: Mount Laguna Observatory, taken from Shafter \etal(1995); DSAZ: Deutsch--Spanisches Astronomisches Zentrum, Calar Alto; AIP: Astrophysikalisches Institut Potsdam)
}{\label{t:obs}}
\begin{center}
\begin{tabular}{|l|l|c|r|r|}
\hline
Date&Telescope&Filter&duration of observation&time resolution\\
\hline
1992/10/30&MLO 1m&V photometry&4:43 h&30 s\\
1992/12/23&MLO 1m&V photometry&3:20 h&31 s\\
1995/11/24&DSAZ 3.5m&spectroscopy&4:56 h&300 s\\
1995/11/26&DSAZ 3.5m&spectroscopy&3:05 h&300 s\\
1995/11/27&DSAZ 3.5m&spectroscopy&7:44 h&300 s\\
1996/08/11 to 1996/10/26&HST/FOS&UV spectroscopy&13 x 12 min&0.8 s\\
1999/11/04&AIP 0.7m&V photometry&2:21 h&30 s\\
1999/11/16&AIP 0.7m&V photometry&3:41 h&30 s\\
\hline
\end{tabular}
\end{center}
\end{table*}

Polars are cataclysmic variables (CVs), in which both stars rotate synchronized with the orbital period. The strong magnetic field of the white dwarf prevents the formation of an accretion disk. Instead it forces the accreted matter - when the magnetic pressure exceeds the hydrodynamic pressure -- to move along the magnetic field lines to one or both magnetic poles (see e.g.~Cropper 1990).

\v13 (\rxj05) was discovered in the ROSAT all-sky survey 
(Beuermann \& Thomas 1993) 
and optically identified as magnetic cataclysmic variable by 
Garnavich \etal(1994) and Walter \etal(1995). 
It is one of the rare systems among the polars, which show a total 
eclipse of the white dwarf.
This gives restrictions which make it much more easy to derive the system parameters from observational data than for most of the other systems.  

\v13 has an orbital period of 7.98 h, by far the longest among all 
known polars. 
The spectral type of the secondary star is between M0 and M1 
(Shafter \etal1995). A main sequence star of this type 
has a much smaller radius than the size of the Roche lobe at the 
given orbital period, i.e.~V 1309 Ori
is the only known polar with a significantly oversized or possibly 
evolved secondary.

The magnetic field of the white dwarf could be infered from 
cyclotron harmonic emission to be $\sim$ 61 MG (Shafter \etal1995).
There was some discussion in the literature, whether with the given 
field strength of the white dwarf and the long orbital period 
the system could rotate synchronously. Frank \etal~(1995) showed that 
\v13 can be understood in terms of the standard evolutionary model,
if the system either was in an extended low state to bring the white 
dwarf into synchronism or that the magnetic field of the secondary is 
strong enough to maintain synchronism.

In the ROSAT light curves large short-term variations of the X-ray flux 
are seen attributed to the impact of individual blobs of matter 
on the accretion region (Walter et al.~1995).

Due to its exceptionally long period, \v13 is highly suited 
for a detailed spectroscopic study involving Doppler tomography. 
The long orbital period allows phase-resolved spectroscopy with high spectral resolution.

In this paper we present high resolution spectroscopy of \v13, obtained with the two channel spectrograph TWIN at the 3.5--m telescope of the German--Spanish Astronomical Centre on the Calar Alto in Spain from 1995/11/24 to 1995/11/28.
Furthermore we present photometry taken with the 70--cm telescope at the AIP site.
We also use V-band photometry published by Shafter \etal~(1995) and archived HST UV spectroscopy, which was recently used by Schmidt \& Stockman (2001).

From these data we derive the eclipse ephemeris, system parameters -- inclination, mass ratio and the primary mass -- and investigate the accretion geometry.

\section{Observations and reduction}

\v13 was observed in 1995 at the 3.5--m telescope at Calar Alto with the double--beam spectrograph TWIN, covering wavelength ranges from approx. 4200 \AA~to 5050 \AA~and from 7550 \AA~to 8700 \AA. The spectral resolution of this data is $\sim$2.3 \AA~($\sim$ 84 \kmps~at 8183 \AA) in the red channel and $\sim$ 1.8 \AA~($\sim$ 115 \kmps~at 4686 \AA) in the blue one.

Flat fields, comparison lamp spectra and bias frames were taken regularly. 

For photometric calibration we took spectra of spectrophotometric standard stars. Seeing and transparency were variable during our observations, 
the photometric calibration of the spectra is supposed to have an 
uncertainty of up to a factor 2.

To avoid the long readout times of the CCDs, we put six -- in slit direction shifted -- single spectra onto the chips before reading out. This caused the night sky lines to be very bright in each spectrum, producing large errors at the corresponding wavelengths.

By taking the data of all three nights, full phase coverage is achieved with a phase resolution of 0.0125 (80 phase bins per orbital cycle). \v13 could 
be observed for 6.5 hours subsequently from Calar Alto. At the epoch of our 
observation, eclipses happened just before and just after the phase of 
visibility, hence, eclipse phase is poorly covered by our spectroscopic
observations.

Photometric data in the V-band were obtained by observations at the 70--cm telescope at the AIP in November 1999. The observations with a time resolution of 30 s cover two eclipses. These data were calibrated with bias, flat field and photometric standard star exposures.

In addition we used an archived HST/FOS observation, which was previously published by Schmidt \& Stockman (2001). This is time resolved spectroscopy in the UV with a time resolution of $\sim$ 0.8 s. \v13 was observed 13 times for approx. 12 min., including one ingress and four egresses of the eclipse. By phase--folding the data it is possible to create a composite eclipse light curve, covering the phase interval from approx. -0.1 to 0.1.

All observations used in this paper are listed in Table~\ref{t:obs}.

\FloatBarrier

\section{Results and analysis}

\subsection{Determination of the ephemeris from photometric data (optical \& UV)}
\label{s:sect31}

The knowledge of an accurate value of the binary period and of the inferior conjunction of the secondary star is a necessary prerequisite for any further modeling of the system. 

The long-term baseline for the determination of the period comes from the combination of our own photometry with published photometry from Shafter \etal(1995) and an archival HST observation.

The times of mid-eclipse of the new data were determined using the same approach as Shafter \etal(1995). Straight lines were fitted to the 
ingress and egress parts of the light curves. The time of mid-eclipse was taken as
the centre of the line drawn at half-ingress intensity between the two fitted lines.

\begin{table}[htb]
\small
\caption{The times of mid-eclipse used for the determination of the ephemeris of V1309 Ori.
}{\label{t:05hjd0}}
\begin{center}
\begin{tabular}{|l|r|l|r|c|}
\hline
Date&Cycle&time of mid-eclipse&Error&Filter\\
\hline
1992/10/30&$-4250$&2448925.8348&30 s&V\\
1992/12/23&$-4088$&2448979.7193&30 s&V\\
1996/09/13&0&2450339.43652&15 s&UV\\
1999/11/04&3449&2451486.6147&30 s&V\\
1999/11/16&3485&2451498.5890&30 s&V\\
\hline
\end{tabular}
\end{center}
\end{table}

\begin{figure}[htb]
\begin{center}
\begin{minipage}{88mm}
\psfig{figure=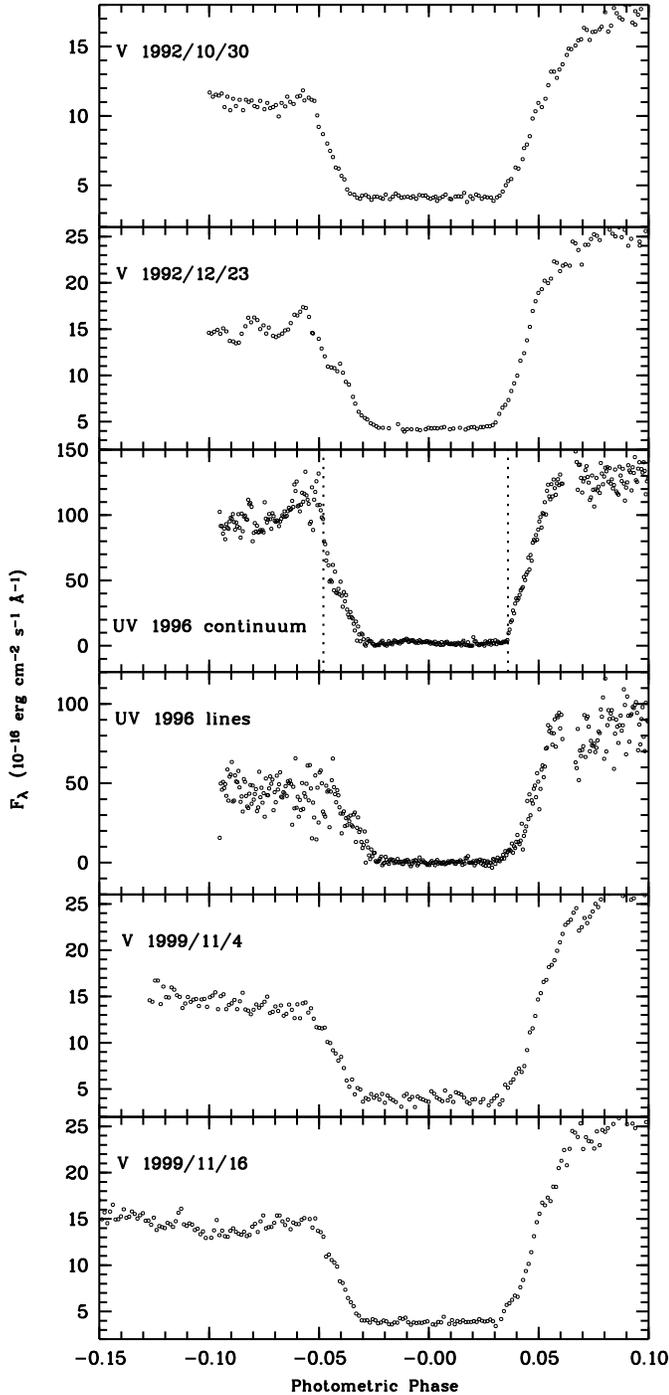,width=88mm,angle=0,clip=}
\caption{\label{f:05eclipses} The V band eclipse light curves used for determination of the ephemeris and the composite UV continuum light curve. In addition, the UV line light curve is shown for comparison. In the UV continuum light curve the ingress and egress of the white dwarf are marked by dotted lines.
}
\end{minipage} 
\end{center}
\end{figure}

By fitting a straight line to these times as a function of cycle number, 
the new photometric ephemeris was derived as
\begin{equation}
\mbox{HJD}(T_{mid-ecl.})=2450339.4363(6)+E*0.33261194(8).
\end{equation}
The numbers in brackets are the errors of the last digits.

The eclipse data used is shown in Fig.~\ref{f:05eclipses}. The phase for all data was calculated according to the ephemeris of Eq. (1). The UV continuum light curve was achieved by averaging the wavelength intervals $1319 - 1361$ {\AA}, $1430 - 1500$ {\AA}, $1576 - 1612$ {\AA}, and $1670 - 1698$ {\AA}. For comparison purposes the UV line light curve, obtained by summing up the intervals $1381 - 1423$ {\AA}, $1531 - 1569$ {\AA}, and $1624 - 1659$ {\AA} and subtracting the interpolated continuum flux, is also shown.

\begin{figure}[htb]
\begin{center}
\begin{minipage}{88mm}
\psfig{figure=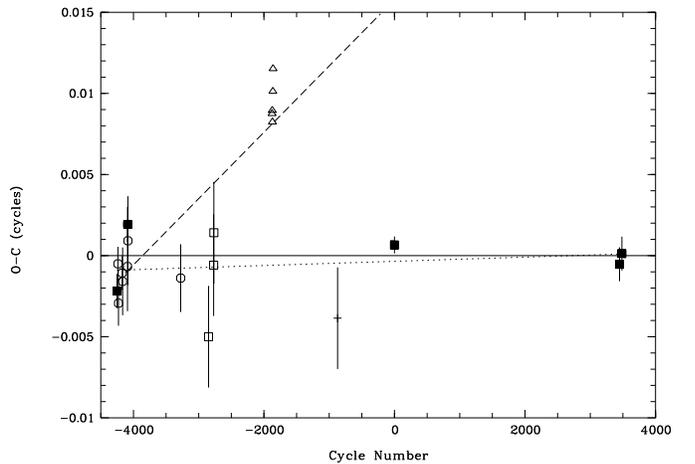,width=88mm,angle=-90,clip=}
\caption{The deviation of observed to calculated mid--eclipse times. Here the filled squares represent the data points used for our period determination (see Table~\ref{t:05hjd0}). The circles denote the eclipse times from Shafter \etal(1995), the empty squares the values published by Garnavich \etal(1994) and the triangles the values published by Buckley \& Shafter (1995). The cross marks the zero point of the spectroscopic ephemeris, derived from the velocity zero crossing of the NaI lines in our spectroscopic data. The dotted line symbolizes the ephemeris from Shafter \etal(1995) and the dashed line the one by Buckley \& Shafter (1995).
}
\label{f:05ephomc}
\end{minipage} 
\end{center}
\end{figure}

The deviations of the observed eclipse times from the calculated ones are shown in Fig.~\ref{f:05ephomc}.

As one can see, the orbital period from Eq. (1) is quite similar to the one of Shafter \etal(1995). Our time base, however, is much larger than the one used there, making our period determination much more accurate.

The ephemeris from Buckley \& Shafter (1995) (indicated by the dashed line in Fig.~\ref{f:05ephomc}) predicts incorrect eclipse times for observations after 1995.

Our spectroscopic observations allow us to determine a single value for the inferior conjunction from the radial velocity of photospheric absorption lines. It is marked with a cross in Fig.~\ref{f:05ephomc} and is consistent with the determined period.

The photometry presented by Shafter \etal(1995) showed the eclipse length to be highly variable, which means that not the white dwarf itself but the accretion stream contributes most to the observed flux.

\begin{figure}[htb]
\begin{center}
\psfig{figure=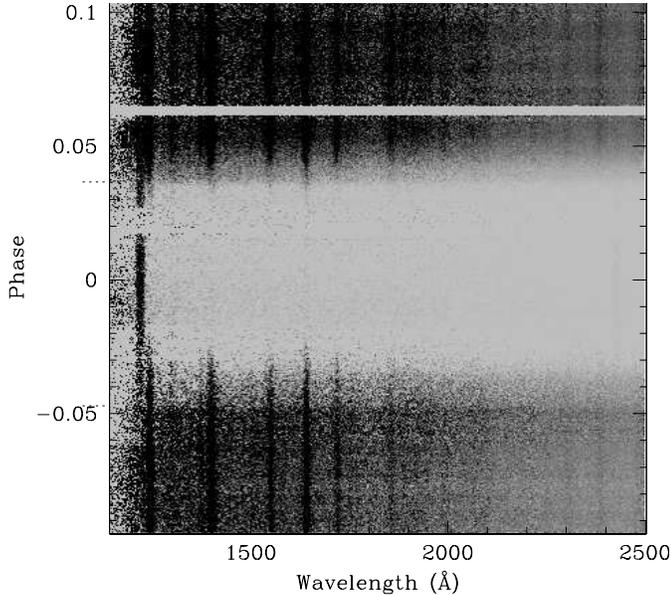,width=88mm,bbllx=40pt,bblly=190pt,bburx=473pt,bbury=582pt,clip=}
\caption{The UV trailed spectrum: The ingress and the egress of the white dwarf are clearly seen at the phases marked with small lines (-0.048, resp. 0.036). The phase is calculated from Eq. (1).
}
\label{f:uvtrail}
\end{center}
\end{figure}

The only data set, which allows us to identify the eclipse of the white dwarf, is the HST observation, which we show as a trailed spectrogram in Fig. \ref{f:uvtrail}.
A UV continuum light curve based on these observations is shown in Fig. \ref{f:05eclipses} (time resolution: 15 s).
The data suggest that (1) the initial steep decline at phase --0.048 is caused by the eclipse of the white dwarf, and (2) that the eclipse egress of the white dwarf marks the end of the phase of totality (phase 0.036).

The analysis by Schmidt \& Stockman (2001) confirms this conclusion, but they claim the detection of a steep decline of length $6 \pm 2$ s, which they attribute to the eclipse of a small hot ($> 150,000$\,K) spot.
Our analysis does not reproduce their results, we see the decline not only in the far UV but also in the near UV. We, therefore, attribute the decline to the eclipse of the whole white dwarf and consequently derive a much lower temperature for the structure which is eclipsed.
Remaining line and continuum emission seen while the white dwarf 
is in eclipse, $\phi = -0.048 \dots -0.030$, indicates that the 
accretion stream is not completely obscured during this phase interval. 

\begin{figure}[htb]
\begin{center}
\begin{minipage}{88mm}
\psfig{figure=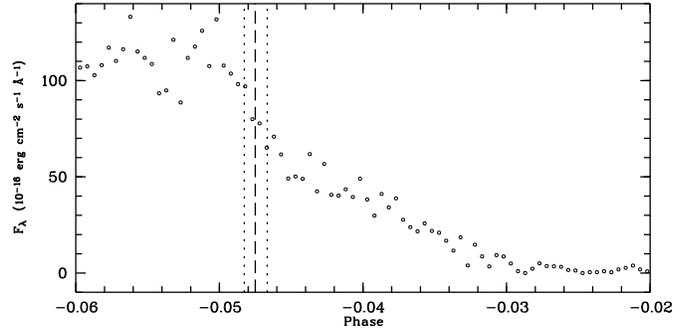,width=88mm,angle=-90,clip=}
\caption{The eclipse ingress in the UV continuum light curve with a time resolution of 15 s. The points of begin and end of the white dwarf's ingress are marked by dotted lines, while phase --0.0475 (dashed line) is equivalent to the center of the ingress. The phase is calculated from Eq. (1).
}
\label{f:uvlc}
\end{minipage} 
\end{center}
\end{figure}

The ingress of the white dwarf eclipse 
lasts $45 \pm 30$\,s (see Fig. \ref{f:uvlc}), which is near the expected value of 71\,s, if we 
calculate it with the system parameters derived in chapter \ref{s:sect33} 
and the mass-radius-relation for white dwarfs from Nauenberg (1972). 

An eclipse duration of the white dwarf 
of 2418 $\pm$ 60 s was determined ($\sim$ 0.084 phase units).
Figs. \ref{f:uvtrail} and \ref{f:05eclipses} show, that there is already some line emission $\sim$ 0.005 phases ($\sim$ 2 min) before the egress of the continuum light source. 

Eclipse ingress and egress of the white dwarf cannot be resolved in 
 broad band optical photometry. This is due to the relative faintness
of the white dwarf in that band -- in Sect.~\ref{s:sect34} we 
estimate its contribution to only about 10\% of the eclipse flux --
and the rather poor time resolution of these observations of only 30 sec.

As one can see in the UV continuum light curve in Fig.~\ref{f:05eclipses}, the center of the eclipse of the white dwarf is not at phase zero, but approx.~0.006 phase units earlier. For conversion from photometric ephemeris to mid-eclipse of the white dwarf one has to subtract 172 ($\pm$ 20) s from the time of mid-eclipse, so that the ephemeris of the white dwarf's superior conjunction is 
\begin{equation}
HJD(T_{wd})=24450339.4343(8)+E*0.33261194(8).
\end{equation}

\FloatBarrier

\subsection{High resolution spectroscopy}

\begin{figure*}
\begin{center}
\resizebox{160mm}{!}{\psfig{figure=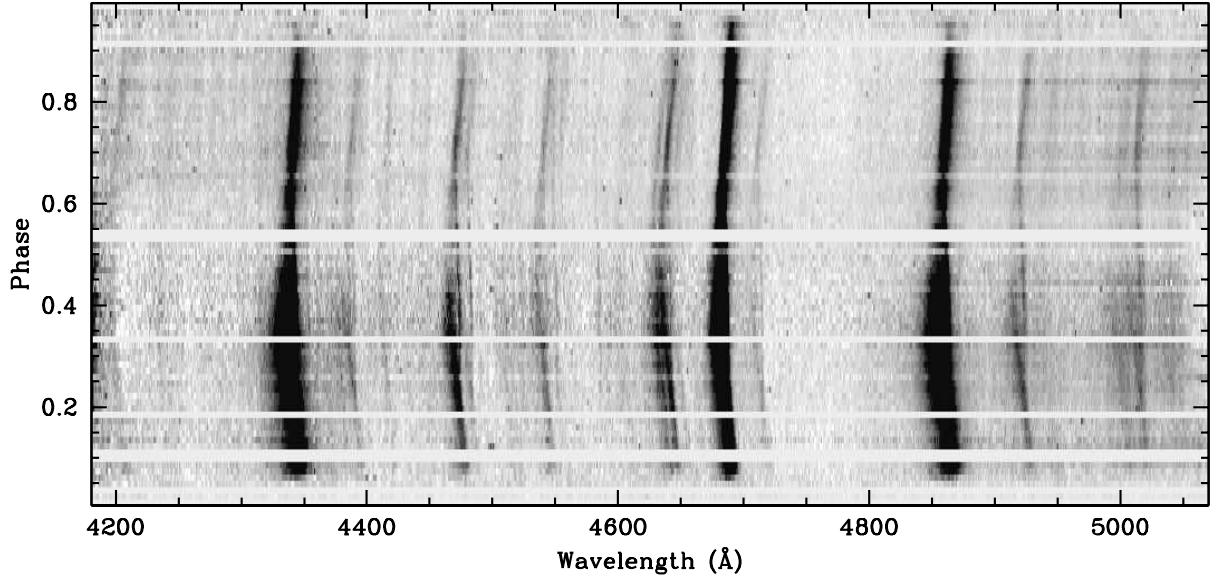,width=120mm,angle=0,clip=}}
\caption{Trailed continuum-subtracted blue spectrum of \v13, created by phase-averaging the individual spectra of all three nights.
}
\label{f:bluetrail}
\end{center}
\end{figure*}

\begin{figure*}
\begin{center}
\resizebox{140mm}{!}{\psfig{figure=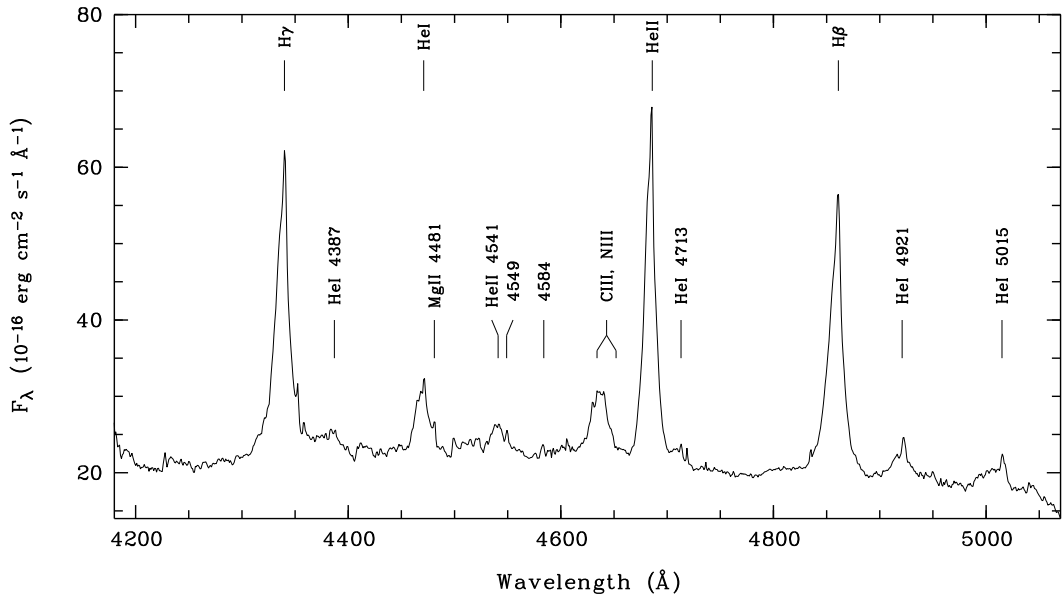,width=120mm,angle=-90,clip=}}
\caption{Mean-orbital spectrum of V1309 Ori in the blue channel, after correction to zero velocity (narrow emission line of \heII).
}
\label{f:bluemean}
\end{center}
\end{figure*}

A phase-binned trailed spectrum of the blue channel is reproduced in Fig. \ref{f:bluetrail}. It is clearly seen that the emission lines have different line shapes and intensities at different phases. It is possible to identify several components in the emission line. One can see that in the emission lines (e.g. \heII\ and H$\beta$ in Fig. \ref{f:he2hbetanatrail}) a narrow emission line component (NEL), nearly following a sine curve, and some more diffuse emission is existent. The NEL is supposed to originate from the irradiated hemisphere of the secondary star (see e.g. Schwope \etal1997), while the other emission is likely to be originating in the accretion stream.

In Fig. \ref{f:bluemean} we show the mean-orbital radial-velocity-corrected spectrum of V1309 Ori in the blue channel of the TWIN. For radial velocity correction we used a sine-fit to the NEL of \heII.

In this spectrum several lines could be identified: the Balmer lines H$\beta$ and H$\gamma$, the lines of neutral (4387, 4471, 4713, 4921 and 5015 \AA) and ionized (4541 and 4686 \AA) helium, the line of Mg\,{\sc ii}\,$\lambda$4481, the CIII/NIII-complex at 4635-4650 \AA~and two unidentified lines at $\sim$4549 and $\sim$4584 \AA. These unidentified lines are interesting, because in the trailed spectrogram (Fig. \ref{f:bluetrail}) they show only the NEL component.

\begin{figure*}
\begin{center}
\resizebox{160mm}{!}{\psfig{figure=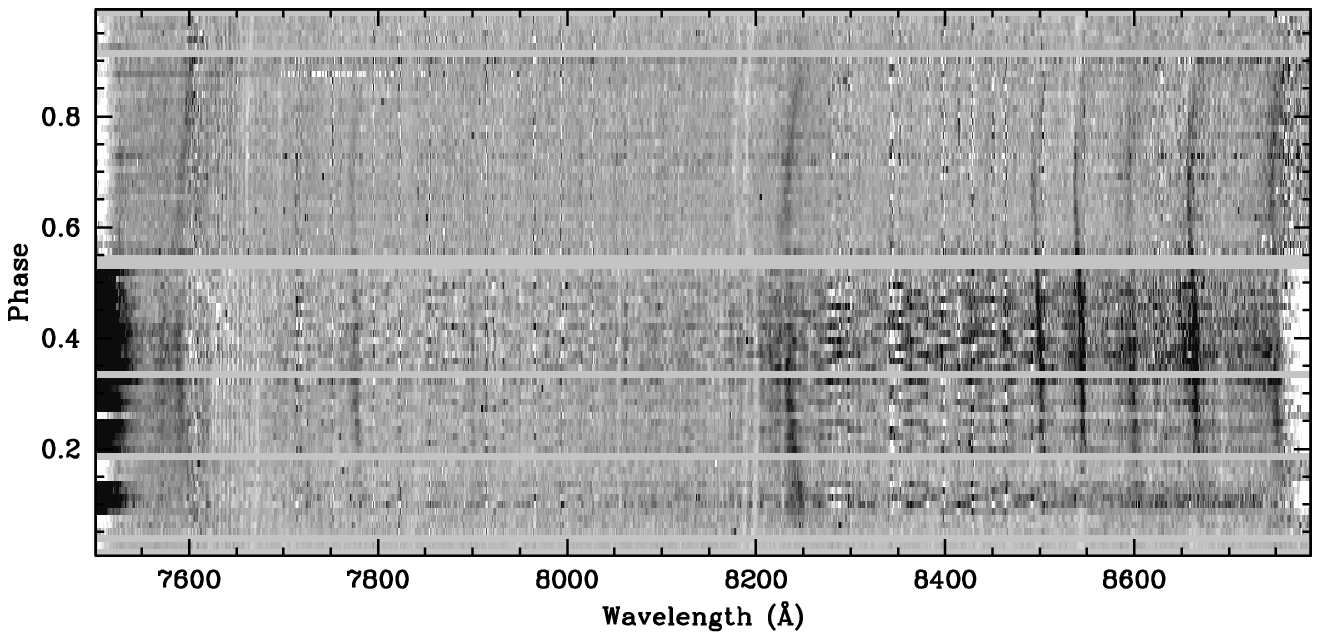,width=120mm,angle=0,clip=}}
\caption{Trailed continuum-subtracted red spectrum of \v13, created by phase-averaging the individual spectra of all three nights. Emission features appear black, absorption features white.
}
\label{f:redtrail}
\end{center}
\end{figure*}

\begin{figure*}
\begin{center}
\resizebox{140mm}{!}{\psfig{figure=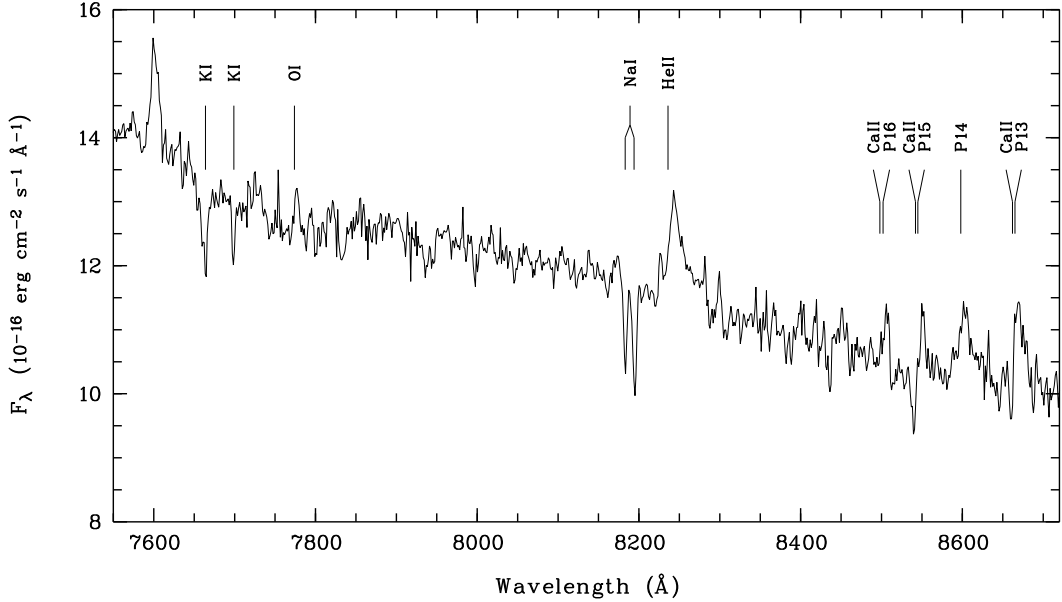,angle=-90,clip=}}
\caption{Mean spectrum of V1309 Ori in the red channel, after correction to zero velocity (NaI absorption lines) and averaging over the phase interval from 0.7 to 1.0.
}
\label{f:redmean}
\end{center}
\end{figure*}

Fig. \ref{f:redtrail} shows the phase-binned trailed spectrogram in the red channel. Some absorption and emission lines are evident while moving back and forth in wavelength.
The area around the NaI doublet at 8183/94 \AA~is magnified in Fig. \ref{f:he2hbetanatrail}.

The spectrum shown in Fig. \ref{f:redmean} was achieved by averaging the velocity-corrected (with a sine-fit to the NaI-lines) red spectra in the phase interval from 0.7 to 1.0.

Several absorption lines (Na\,{\sc i}\,$\lambda\lambda$8183/8194, K\,{\sc i}\,$\lambda$7664 and K\,{\sc i}\,$\lambda$7699) and emission lines (He\,{\sc ii}\,$\lambda$8236, O\,{\sc i}\,$\lambda$7775 and the blend of CaII and Paschen lines above 8400 \AA) could be identified.

\begin{figure*}
\begin{minipage}{60mm}
\psfig{figure=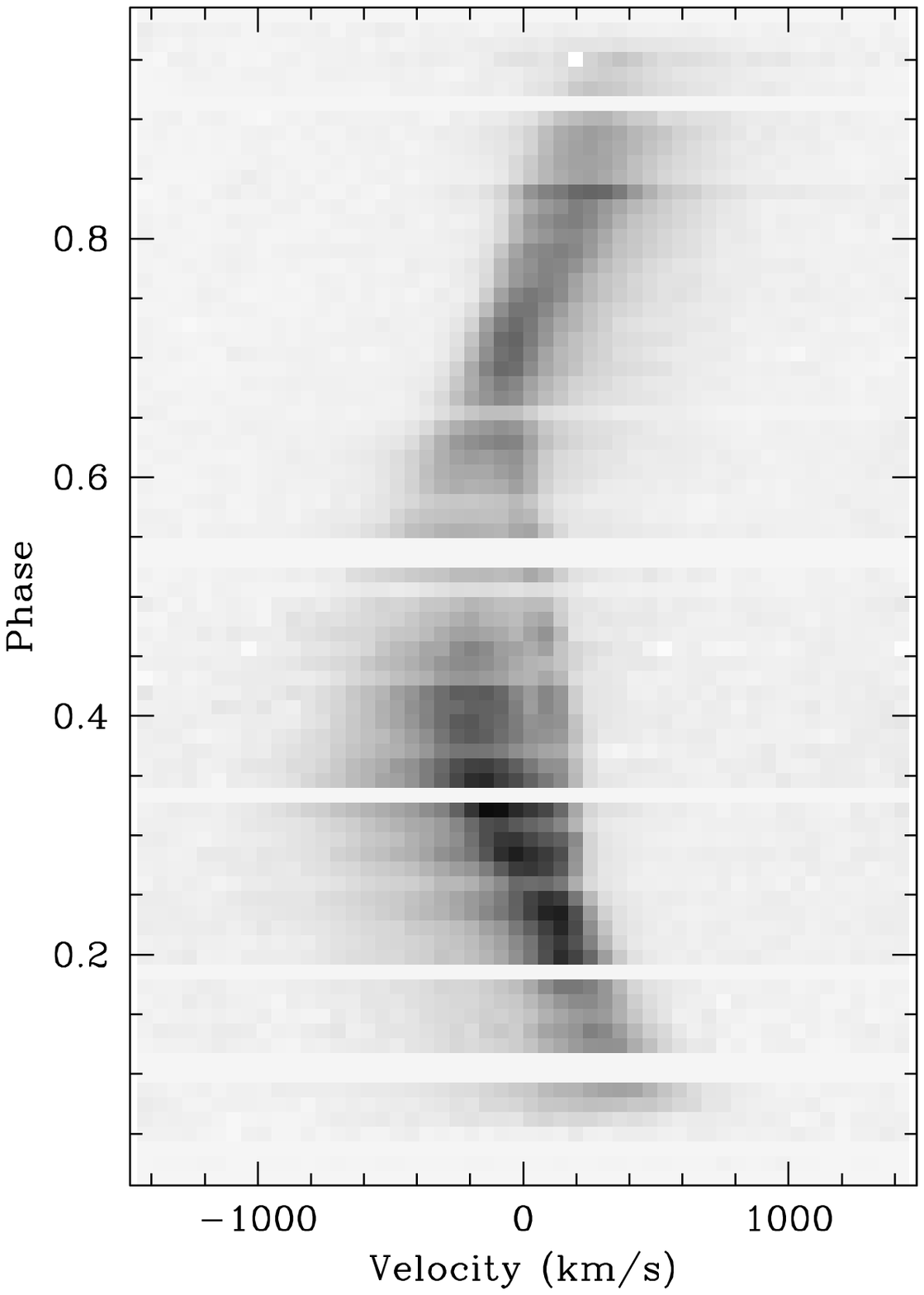,width=60mm,angle=0,clip=}
\end{minipage}\hfill 
\begin{minipage}{60mm}
\psfig{figure=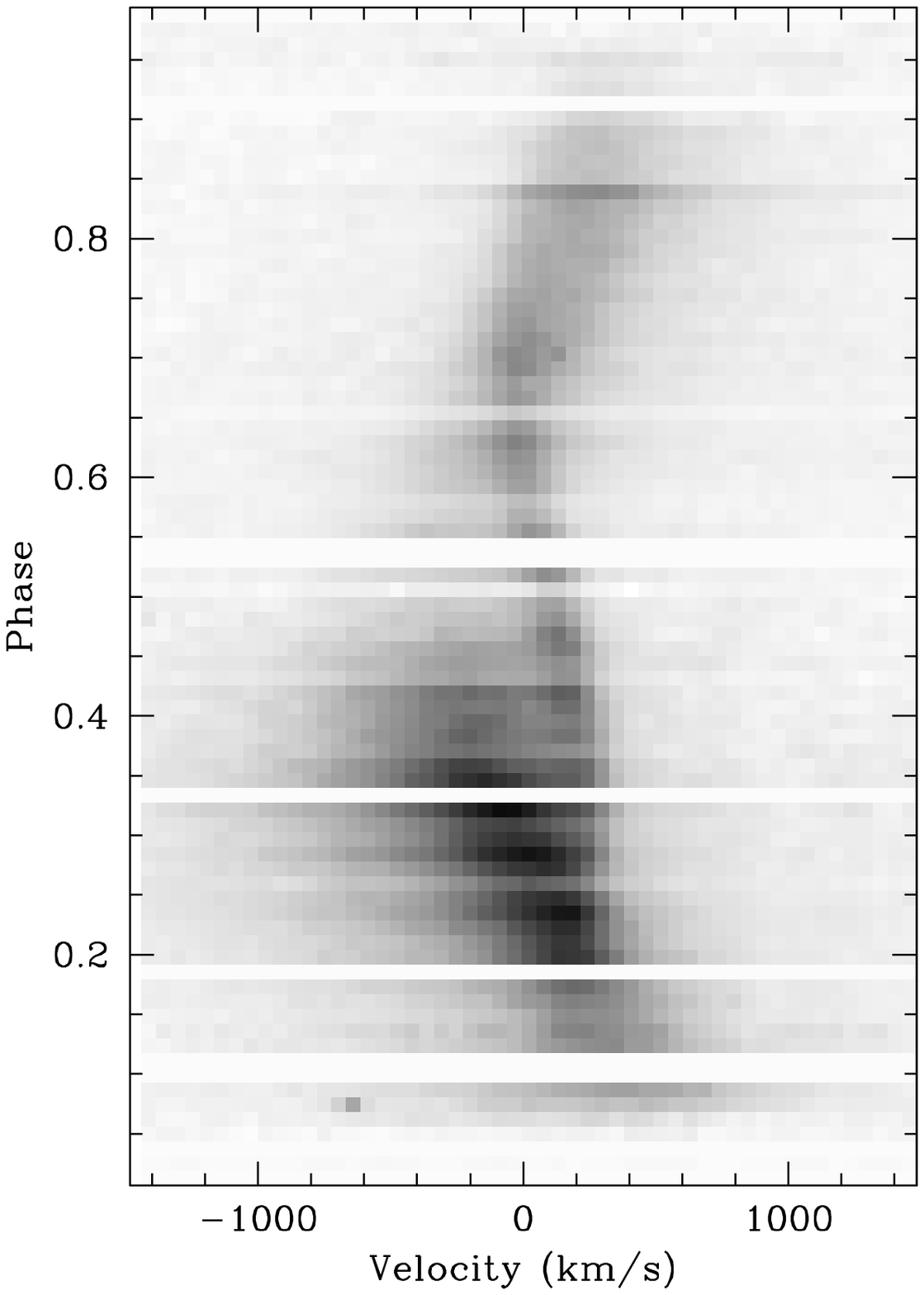,width=60mm,angle=0,clip=}
\end{minipage}\hfill
\begin{minipage}{60mm}
\psfig{figure=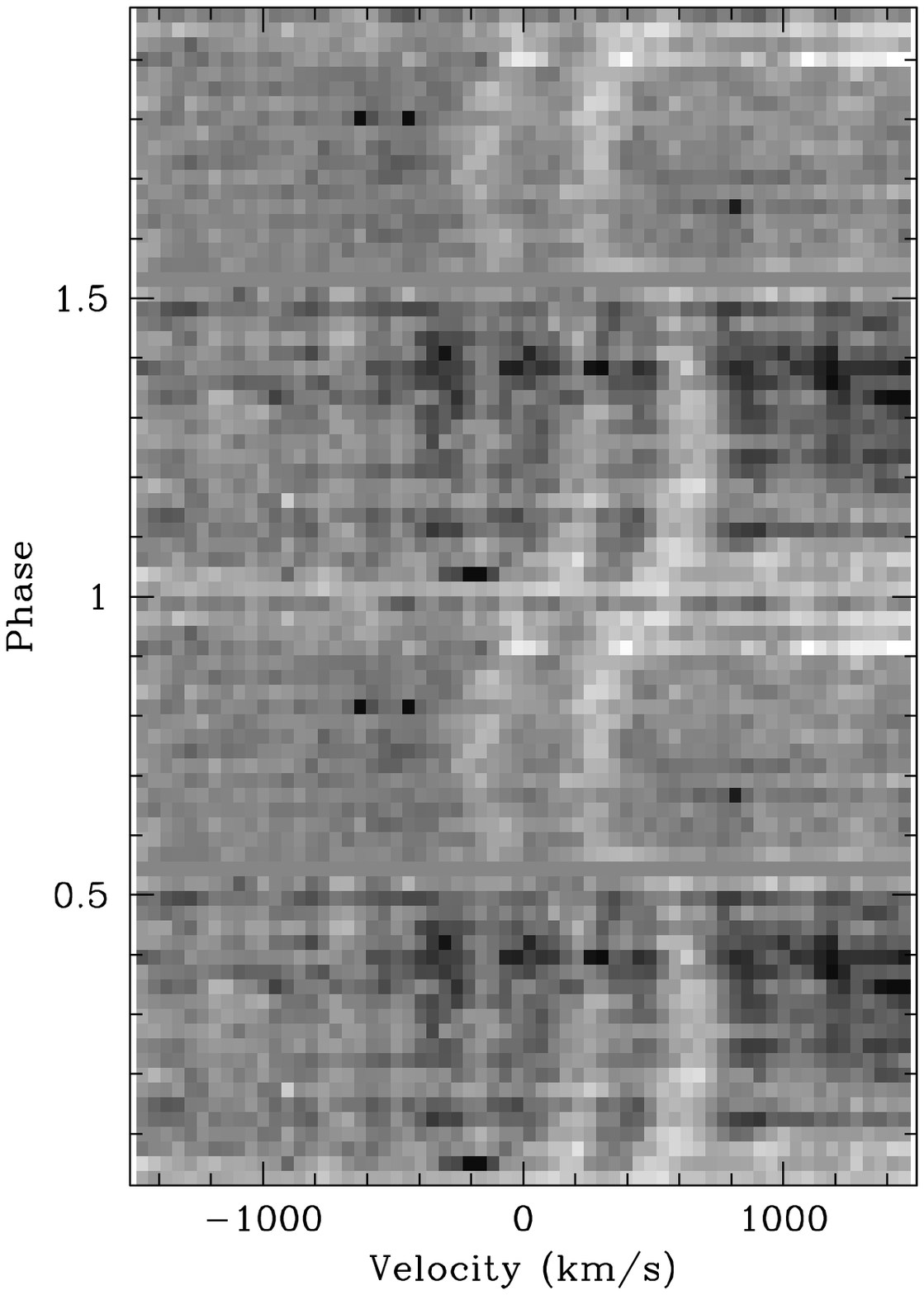,width=60mm,angle=0,clip=}
\end{minipage}\hfill
\caption{Trailed spectra around \heII, around H$\beta$ and around the NaI doublet at 8183/8194 \AA. In case of the latter the velocity is computed for the line at 8183 \AA. For clarity the data of NaI are shown twice.
}
\label{f:he2hbetanatrail}
\end{figure*}

For further analysis it was necessary to determine the system velocity and to remove it from the radial velocity curves and trailed spectra. This was done by fitting two gaussians with fixed separation and same intensity to the NaI lines in each of 40 phase-binned spectra.

By fitting a sine-curve to the centres of the gaussians as a function of
the orbital phase, a system velocity of 8(6) \kmps\ was determined. 
With the 
baryocentric velocity correction of another 8\,\kmps\ the (low) 
system velocity with respect to Earth of $\gamma = 16 (6)$\,\kmps\ follows. 
The value of 8\,\kmps\ was used to correct the observed spectrograms 
before tomographic analysis.

\subsection{The system parameters of \v13}
\label{s:sect33}

In this section we make use of the HST eclipse observation and radial 
velocity measurements of the secondary star in order to determine the 
orbital inclination $i$, the mass ratio $Q = M_{\rm wd}/M_{2}$, and the 
mass of the white dwarf $M_{\rm wd}$. 

\begin{figure}[htb]
\begin{center}
\begin{minipage}{88mm}
\psfig{figure=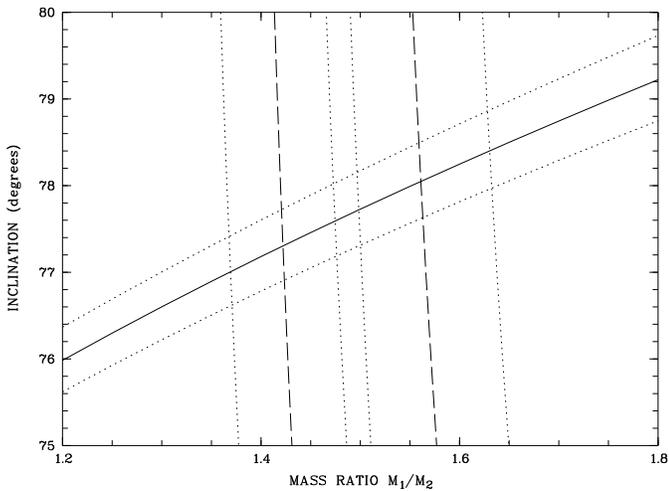,width=88mm,angle=-90,clip=}
\caption{For the measured eclipse duration of 2418 s the curve of possible combinations of inclination and mass ratio is shown (solid line). The dashed lines represent the possible combinations of mass ratio and inclination derived from the radial velocities of the NEL in our irradiation model for $M_{2} = 0.4$\,\msun\ (right) and 0.6\,\msun (left). The dotted lines correspond to the uncertainties of the used parameters, e.g. the nearly vertical lines to the range for $K_2$ and the other ones to the uncertainty of the eclipse duration.
}
\label{f:q_i_diagr}
\end{minipage} 
\end{center}
\end{figure}

As described in Sect.~\ref{s:sect31}, the observed eclipse length of 
the white dwarf in the ultraviolet spectral range is 
$\Delta t = 2418 \pm 60$\,sec. 
For further analysis we assume, that the secondary star fills its Roche 
volume and that the white dwarf may be approximated as a point source.
Then, according to Chanan et al.~(1976),  a purely geometrically determined
relation between $i$, $Q$, and $\Delta \phi$ exists, which is shown in 
Fig.~\ref{f:q_i_diagr}.

\begin{figure}[htb]
\begin{minipage}{43mm}
\psfig{figure=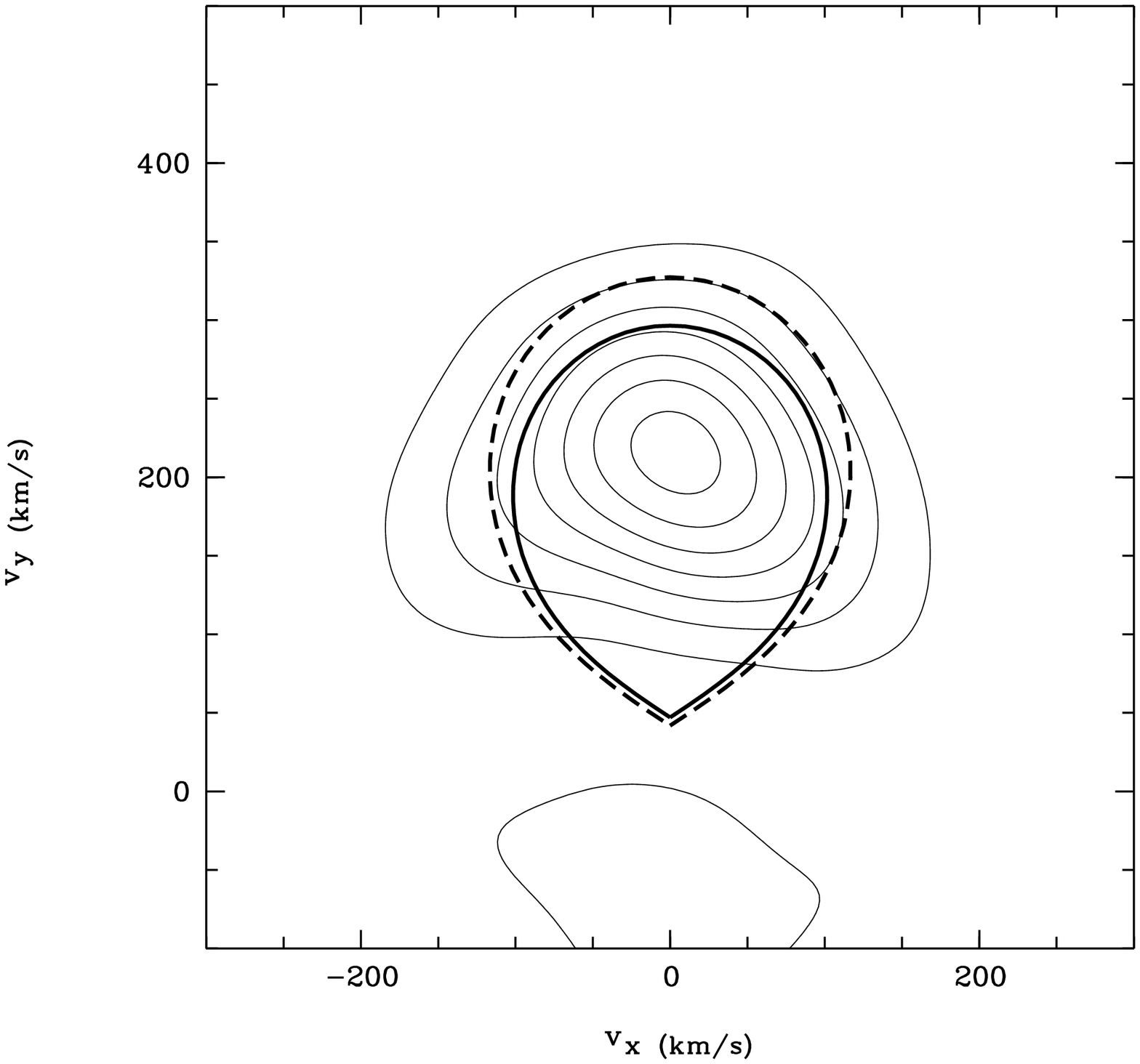,width=43mm,angle=-90,clip=}
\end{minipage}\hfill 
\begin{minipage}{43mm}
\psfig{figure=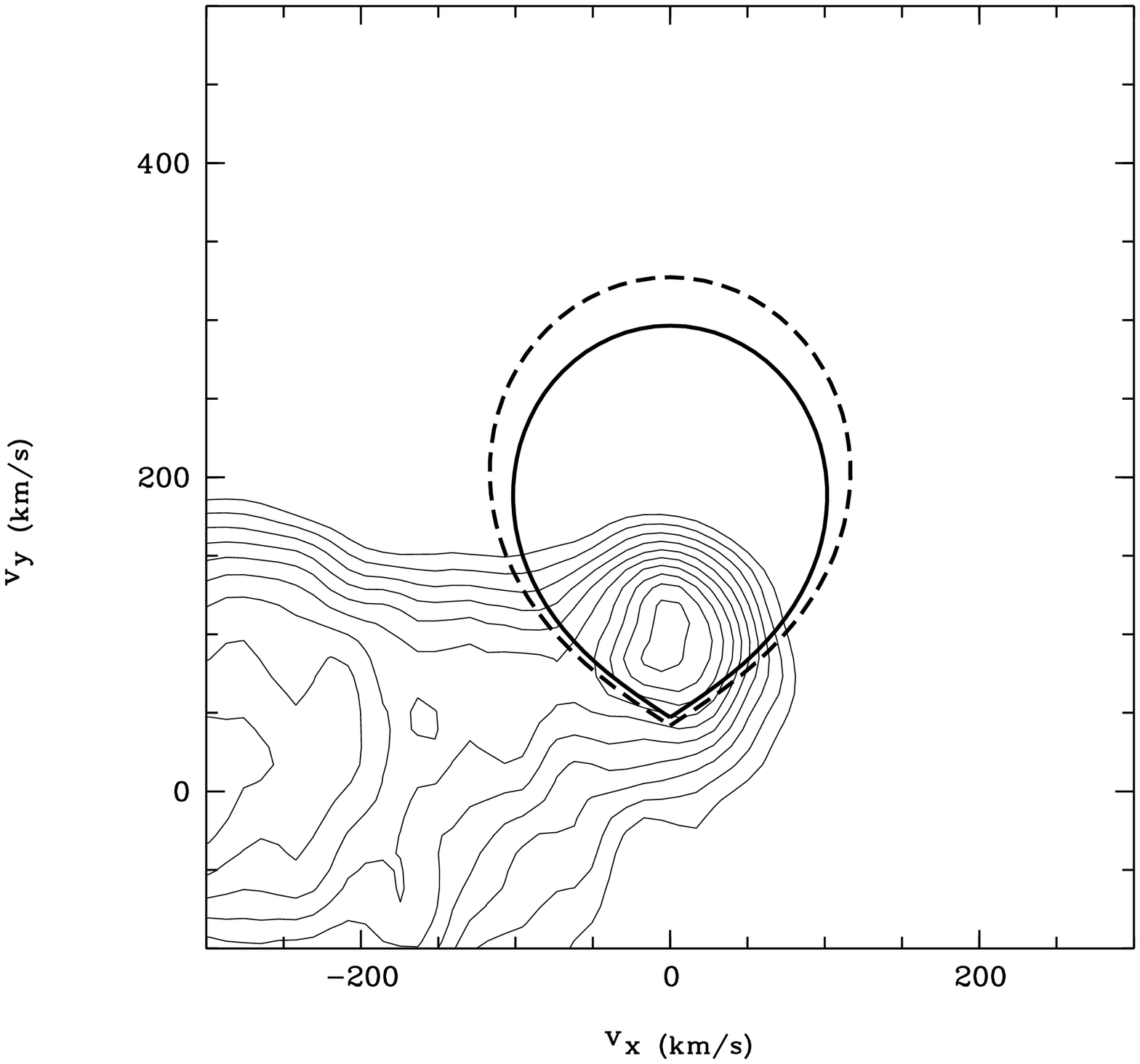,width=43mm,angle=-90,clip=}
\end{minipage} 
\caption{The tomograms of NaI (left) and \heII (right) with the outlines of the Roche lobes for $Q = 1.42$, $M_2 = 0.6$\,\msun, $i = 77.3\degr$ (dashed) and for $Q = 1.56$, $M_2 = 0.4$\msun, $i = 78.0\degr$ (solid). The contour lines are at 10\% steps.
}
\label{f:rochefit}
\end{figure}

The parameters $Q$ and $i$ can further be constrained, if the orbital 
velocity $K_2$ of the secondary star is known. Our spectroscopic data
allow an indirect measurement of this quantity by observations of 
emission and absorption lines originating on the secondary star.
According to their phasing and their photometric variation, the 
prominent narrow emission lines (NEL) must originate on the X-ray irradiated
side of the secondary star. In principle, its radial velocity variation 
gives clues to the orbital velocity of the secondary. The determination 
of the radial velocity amplitude is, due to severe line blending with
components originating from the accretion stream, not done straight 
forwardedly. Recent experience gained from a study of another 
long-period polar, QQ~Vul (Schwope et al.~2000), revealed Doppler 
tomography as appropriate tool to determine the NEL radial velocity.

Doppler maps using Spruit's (1998) MEM code were computed for the 
principal emission lines H$\gamma$, H$\beta$, and \heII.
The Doppler map of \heII\ is reproduced as contour plot in 
Fig.~\ref{f:rochefit}. The radial velocity was determined from 
Gaussian fits to an average profile along the $v_{y}$-axis. 
All NEL-velocities agreed within $5$\,\kmps,
suggesting a common location of emission. The velocity amplitude 
used for further analysis is $K_{2}^{\prime} = 89$\,\kmps.

Also, the radial velocity variation of the Na-absorption lines
can be used to constrain the orbital velocity of the secondary star.
The trailed spectrogram of Fig.~\ref{f:he2hbetanatrail} immediately shows that
the Na-lines have a much larger radial velocity amplitude than 
the narrow emission lines, that the radial velocity curve deviates
from a sinusoid, and that the lines almost disappear at phase $\phi = 0.5$, 
when the irradiated side of that star is best visible. Altogether
this means that the line formation and the radial velocity curve 
is strongly influenced by irradiation from the accretion region on the
white dwarf. This becomes obvious from a Doppler map of the 
line doublet also shown in Fig.~\ref{f:rochefit}. The region near
the inner Lagrangian point $L_{1}$ is completely devoid of Na-line 
absorption.

Radial velocities of Na-lines in individual spectra were measured
by fitting a double Gaussian. The radial velocity amplitude of the 
lines were determined by fitting circular and elliptical orbit equations to 
the measured phase-dependent velocities, we consistently 
arrived at $K^{\prime\prime}_{2} = 216\pm7$\,\kmps.

Using the measurements of the radial velocity amplitudes of the photocenters
$K^{\prime}_2$ and $K^{\prime\prime}_2$ of the irradiated and non-irradiated 
hemispheres of the secondary, the orbital velocity can be calculated
with an irradiation model. The details of the model were described 
just recently by Schwope et al.~(2000) and need not be repeated here.
It assumes that emission lines are formed at the irradiated part
of the secondary's Roche surface and absorption lines only on the
complementary region. For given period, inclination, mass ratio 
and mass of the secondary factors for correction from centre of mass to 
centre of light result. 

The correction factor for the velocities of the photocenters of the 
non-irradiated hemisphere to the center of mass, $K^{\prime\prime}_{2}/K_{2}$, 
is only slightly dependent on the mass ratio. It ranges from 1.13
for $Q = 2.5$ to 1.18 for $Q=1.25$. Hence, with the observed radial 
velocity amplitude of the Na-lines and the maximum correction factor
the lowest orbital velocity of the secondary star is $K_{2} = 180$\,\kmps.
The resolution of the Doppler map of the Na-lines is not sufficiently 
high in order to decide whether the irradiated part of the secondary 
is at least partially visible in the Na-lines. We, therefore, allow also
for a smaller correction factor $K^{\prime\prime}_{2}/K_{2} = 1.08$. 
Our accepted range for the orbital velocity thus becomes 
$K_2 = 180 - 200$\,\kmps.

In order to determine the mass of the secondary star in polars, one usually
assumes that it is a main-sequence star. The size of the Roche-lobe is fixed 
for a given binary period. The mass of the secondary star can then be
determined with the aid of a mass-radius relation for main-sequence stars.
As pointed out by Garnavich et al.~(1994) and Shafter et al.~(1995), 
the secondary is of spectral type M0--M1 and, hence, much less massive then
a main-sequence star which fits in a Roche lobe at 8 hours orbital 
period (which would have about 1\,\msun). 
The secondary star must have a mass as low
as 0.6\,\msun\ or less. Assuming $M_{2} = 0.6$\,\msun, Kepler's third law gives
$Q=1.16 - 1.38$, and $M_{\rm wd} = 0.70 - 0.83$\,\msun\ for the allowed 
range of $K_{2}$. If we assume $M_2 = 0.4$\,\msun, we get $Q=1.44 - 1.73$, and
$M_{\rm wd} = 0.57 - 0.69$\,\msun.

We also computed a grid of models for the NEL originating on the irradiated
hemisphere of the secondary star as a function of $i$, $Q$, and $M_2$.
The results are shown as iso-velocity lines for $K^{\prime} = 89\pm5$\,\kmps\
for $M_{2} = 0.4$\,\msun\ and 0.6\,\msun, respectively, in Fig.~\ref{f:q_i_diagr}.
These lines cross the lines defined by the eclipse length almost orthogonally 
and give narrow constraints to the mass ratio and inclination. The accepted
ranges are $i=76.6\degr - 78.9\degr$, $Q = 1.37 - 1.63$, 
$M_{\rm wd} = 0.82 - 0.58$\,\msun, and $M_2 = 0.6 - 0.4$\,\msun. 
The outlines of the Roche lobes for two combinations (the ones defined by the intersections of the dashed lines with the solid one in Fig.~\ref{f:q_i_diagr}) are plotted over the tomograms of sodium and \heII. The areas of origin of the spectral lines lie well within the Roche lobes.
In principle, the size of the Roche lobe (i.e.~$Q$) 
is determined by the ratio $K^{\prime\prime}_{2}/K^{\prime}_{2} = v_b/v_f$. 
A corresponding graph for $v_b/v_f$ vs.~$Q$ 
is shown in Schwope et al.~(2000, Fig.~10).
Our measured ratio $K_{\rm Na}/K_{\rm NEL} = 216/89 = 2.4$ 
predicts a mass ratio $Q=1.63$, at the 
edge of the range given above. We, therefore, tend to accept a solution 
with a larger mass ratio and a smaller secondary mass.
For further modelling the combination $Q=1.5$, $M_{\rm wd} = 0.70$\,\msun\
 was used.

\subsection{Distance estimation}

The observed V-band flux of \v13 during the eclipse is 17\fm45(10) (AIP, 1999/11/16) 
(in full agreement with $V_{\rm ecl} = 17\fm37$ in Shafter \etal1995).
We use YY Gem (A) as reference M0.5-star for an estimation of the distance. 
In Chabrier \& Baraffe (1995) the absolute V-band brightness, the distance and the radius of this star are listed {\bf{($M_V = 8\fm99$, $R = 0.66 \pm 0.02$ \rsun, $D = 13.89$ pc)}.}

Warner (1995) gives a formula for the Roche lobe equatorial radius $R_L ({\rm eq})$
(in the orbital 
plane) as a function of the mass ratio $Q$ and the binary separtion $a$.
For the most likely values in our parameter determination, $Q = 1.5$ 
and $a = 1.48 \times 10^{9}$m (for $M_2 = 0.46$), we calculate $R_L ({\rm eq})
= 5.15 \times 10^{8}$m.
This radius is 12\% larger than the one of YY Gem (A) ($4.59 \times 10^{8}$m),
 which supports the supposition of a larger radius of the secondary in \v13 than of a main sequence star with the same mass.
Both stars have the same spectral type, i.e.~same
surface brightness. Scaling the flux of YY Gem (A) to the larger area of \v13, one gets an absolute brightness of $M_V = 8\fm74$. The distance modulus then gives a reasonably good
estimate of the true distance of $D = 550 \pm 75$\,pc. The error includes 
the uncertainties of the mass ratio and of the secondary mass.

Our distance determination has to be compared to those in the literature.
Shafter \etal(1995), by using the same method, arrive at the same distance 
($D \ge 500$\,pc)
as we do. They claimed their distance to be a lower limit due to a possible 
contamination by the white dwarf and the accretion stream during eclipse. 
The HST-spectroscopy shows, that the eclipse of stream and white dwarf are total
so that the distance of 550\,pc can be regarded as {\bf{an}} exact value.

Harrop-Allin \etal(1997) derive a much larger distance of $D>1300$\,pc from 
K-band spectroscopy. This estimate is based on the non-detection of certain 
M-star features in a single spectrum which they interpret as due to the faintness
of the secondary. However, their spectrum was taken during orbital phase 0.43 to 
0.56, i.e.~at phases where irradiation fundamentally changes the spectrum of the 
secondary as is evident from our phase-resolved NaI-spectroscopy. 
Hence, their estimate is not reliable.

A value of $D=745 \pm70$ pc, derived from the TiO-band strength, is given by 
Beuermann (1999). We cannot comment on the likely reason for the discrepant
distance determinations, a possible cause might be non-solar abundances in the
secondary's atmosphere. For further analysis we use $D= 550$\,pc.

\subsection{Temperature of the white dwarf}
\label{s:sect34}

Based on our determinations of the mass $M_{\rm wd}$ and the distance $D$ 
we can estimate the temperature of the white dwarf from HST-spectroscopy. 
Due to low signal-to-noise in the individual HST-spectra it is difficult to 
extract a meaningful spectrum of the white dwarf from the data at ingress and
egress phase. Instead
we used the average flux in the line-free spectral interval between 1430\,\AA\
and 1500\,\AA\ and compare it with model atmosphere spectra for pure hydrogen 
atmospheres kindly provided by B.~G\"{a}nsicke (G\"{o}ttingen). The brightness 
steps at ingress and egress are not equal, that at eclipse egress is smaller. Whether
the combined emission from the undisturbed white dwarf and a hotter accretion spot
or an accretion flare right at eclipse ingress are responsible for this 
difference cannot be answered with the present data due to the limited number 
of photons and due to the fact that this binary phase was covered 
only once with HST. 
In order to arrive at an upper limit for the white dwarfs temperature
we take the flux difference at eclipse egress, which is 
$F_{1430-1500} \le 4 \times 10^{-15}$\,\erga. 
Using $R_{\rm wd} = 7 \times 10^8$\,cm
for an 0.8\,\msun\ white dwarf, which corresponds to the white dwarf 
with the smallest surface within the derived mass range, and $D=550$\,pc 
we estimate $T_{\rm wd} \le 20,000$\,K.

Schmidt \& Stockman (2001) derived a {\it{lower}} limit of the white dwarf's temperature of 30,000 K. However, they made not clear how their value was derived. The reason for the differing results therefore cannot be identified.

With our parameters the contribution of the white dwarf to the V-band 
is less than $3 \times 10^{-17}$\,\erga, which is less than 10\% of the eclipse
flux.

\subsection{The accretion geometry}

Clues to the accretion geometry can be derived from both photometric 
and spectroscopic 
data. In particular we make use of Doppler images of the main 
emission lines
and the contact phases of certain features in emission line and 
continuum light curves.

\begin{figure}[htb]
\begin{center}
\begin{minipage}{88mm}
\psfig{figure=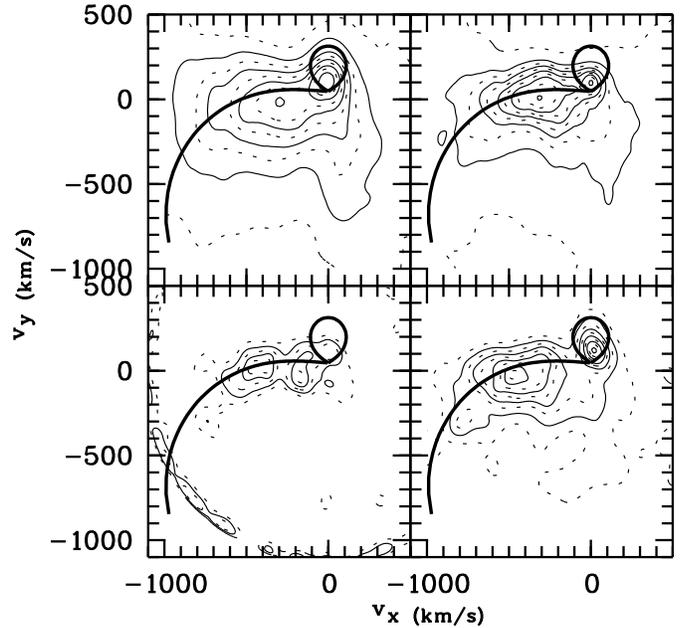,width=88mm,angle=-90,clip=}
\caption{The tomograms of the emission lines: left (from top to bottom): H$\gamma$ and He\,{\sc ii}\,$\lambda$8236, right: \heII~and He\,{\sc i}\,$\lambda$4471. Shown are also the Roche outlines for the above derived system parameters and the line for an assumed ballistic stream.
}
\label{f:bluetom}
\end{minipage} 
\end{center}
\end{figure}

In Fig.~\ref{f:bluetom} we show Doppler images of H$\gamma$, He\,{\sc ii}\,$\lambda$8236, \heII~and He\,{\sc i}\,$\lambda$4471. 
The velocity resolution of the maps is of order 120\,\kmps\ within a radius of 500\,\kmps around the origin. 
That the features in the tomogram of H$\gamma$ appear broader than the ones in the case of \heII~could be caused by blending with He\,{\sc ii\,$\lambda$4339}.
All lines, except He\,{\sc ii}\,$\lambda$8236, clearly show the NEL from the 
irradiated hemisphere of the secondary. Beside the NEL, emission from the 
accretion stream can be recognized in the left and lower quadrants of the
Doppler maps. The overlays show the size of the Roche lobe for our favourite
parameter combination (Sect.~\ref{s:sect33}) and the expected location 
of the ballistic accretion stream. 

In all of our maps emission near the ballistic trajectory is visible. 
In He\,{\sc ii}\,$\lambda$8236 we find emission only in the vicinity of the 
ballistic stream. 
It is possible that the deviation of this tomogram from that of \heII~is caused by the smaller signal-to-noise ratio.
But in each of the other tomograms there is emission off the ballistic trajectory, which can only be due to emission from the magnetic stream.
However, ballistic and magnetic stream do not appear as separate structures in 
the map. It is, therefore, not easy to determine a terminal velocity of
the ballistic stream. We regard the spot of enhanced emission 
at $(v_x,v_y) \simeq (-450,+30)$\,\kmps\ as {\bf{the}} likely region where 
most of the matter couples onto magnetic field lines. 

Our tomograms are very similar to the ones of H$\beta$ and \heII~presented by Hoard (1999), where strong emission from the ballistic stream is seen.
The only major difference is the claimed absence of the NEL in \heII, which is clearly present in our data.
Possible explanations are the higher resolution of our data or a higher accretion state of the system during our observations, which caused a stronger irradiation of the secondary's surface.

Although not identified as separate structure in the maps, some constraints
on emission from the magnetic stream can be set. Firstly, 
emission is spread over a very extended region in Doppler 
coordinates indicating some kind of accretion curtain to be existent.
Secondly, the bulk of emission is concentrated above a line 
$v_y = - 200$\,\kmps, 
which constrains the orientation of the velocity vectors along 
the accretion curtain. The maps differ in this respect clearly from 
those of e.g.~HU Aqr, UZ For, and QQ Vul, which all show the 
magnetic stream to be bright in the lower left quadrant (Schwope 2001).
These three rather well studied systems all have a 'standard' orientation
of the magnetic field, i.e.~the magnetic axis is inclined towards
the ballistic stream. The discrepant appearance of the Doppler maps 
of \v13\ indicates a non-standard orientation of the magnetic axis.

The HST-based emission line light curve shows stream emission up to 
phase $-0.030$, i.e. $0.018$ phase units after the white dwarf is eclipsed
by the secondary. Also, egress of the emission lines is observed
prior to egress of the white dwarf by $0.004$ phase units. Any model 
of the stream geometry has to account for this extended visibility.

The optical $B$ and $V$ band light (see Shafter \etal1995) 
curves did not resolve eclipse
ingress and egress of the white dwarf. 
Outside eclipse they are double humped with 
a primary maximum at binary phase 0.25. 
Around eclipse they show extended ingress and
egress phases due to the eclipse of the accretion stream (or accretion 
streams).  The start phase of totality varies
between $\phi = -0.024 \dots -0.037$. 
The stellar photospheres of the white dwarf primary and the red secondary 
make small contributions to the $BV$ light curves only, i.e.~these
bandpasses are completely dominated by the accretion stream. 
At the given inclination and the extent of the ballistic stream, eclipse
egress of the ballistic stream starts only at phase $\phi = 0.068$, when 
about 75\% of the total light from the stream is visible again after eclipse.
We conclude, that the fraction of light originating from the magnetic
stream/accretion curtain is about 75\%, that from the ballistic stream 
only about 25\%. The pronounced orbital variability finds an 
explanation in projection of an optically thick accretion stream/accretion 
curtain along 
the line of sight. Different 
height of the primary and secondary maxima in the broad-band light 
curves (e.g. $m_B = 15\fm85$ at phase $\phi = 0.25$ and $m_B = 16\fm30$ at phase $\phi = 0.65$ on December 22nd, 1992) are explained by strong irradiation from the white dwarf. 
Obviously, the maximum projected area of the irradiated part of the 
stream/curtain is seen around phase 0.25.
 
We developed a 3D model of the binary which predicts for a given set 
of binary parameters  ($i, Q, K_1, K_2$), for an assumed accretion 
geometry and for an assumed magnetic field strength the location of ballistic 
trajectories in spatial and Doppler coordinates, schematic Doppler
tomograms, and light curves. For the generation of the
light curves we assumed constant emissivity along the 
stream between $L_1$ and the white dwarf. Hence, we cannot expect a 
proper fit to oberved data, but reproduce observed contact phases.

\begin{figure}[htb]
\begin{minipage}{43mm}
\psfig{figure=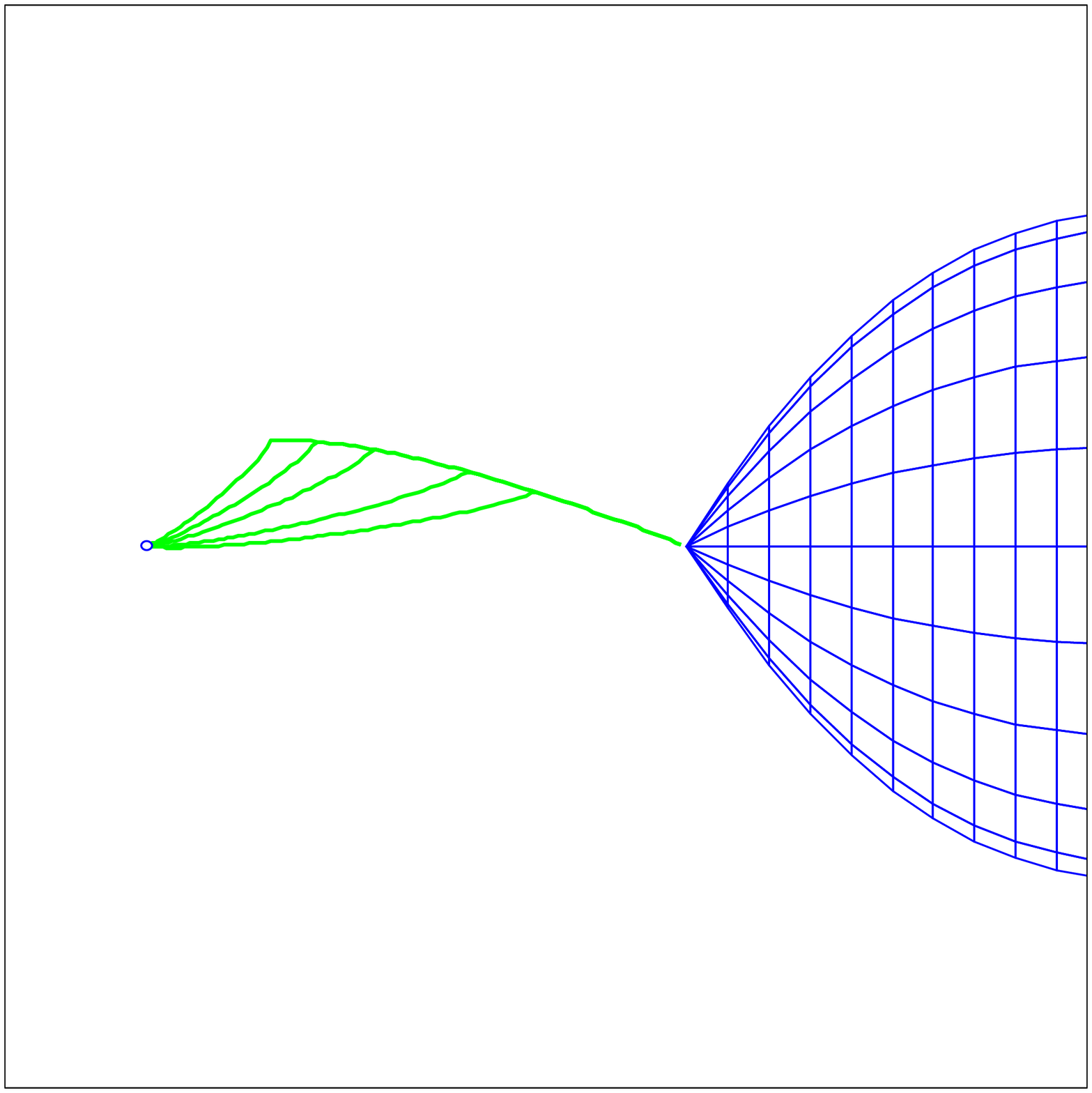,width=43mm,angle=0,clip=}
\end{minipage}\hfill 
\begin{minipage}{43mm}
\psfig{figure=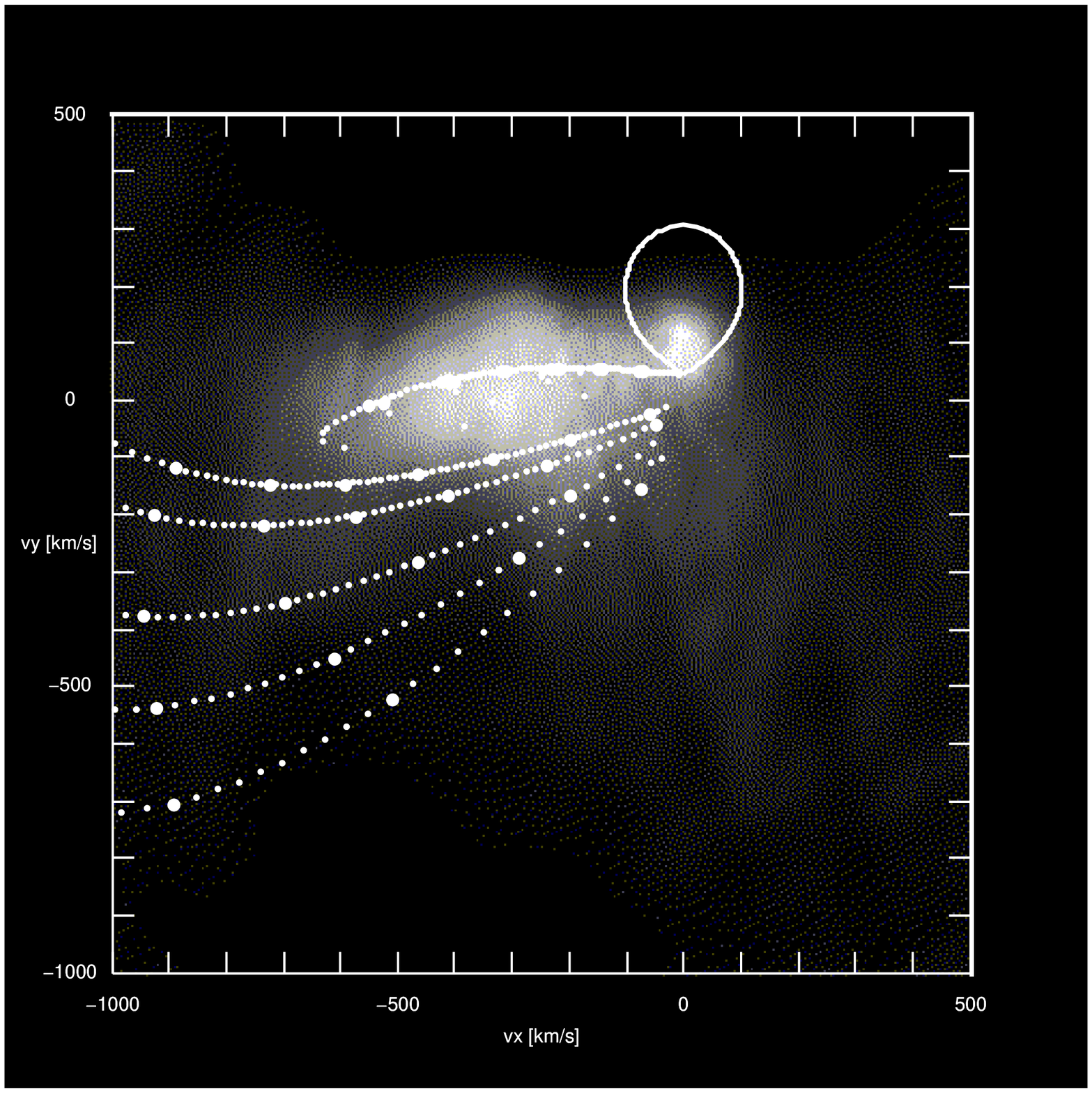,width=43mm,angle=0,clip=}
\end{minipage} 
\begin{minipage}{43mm}
\psfig{figure=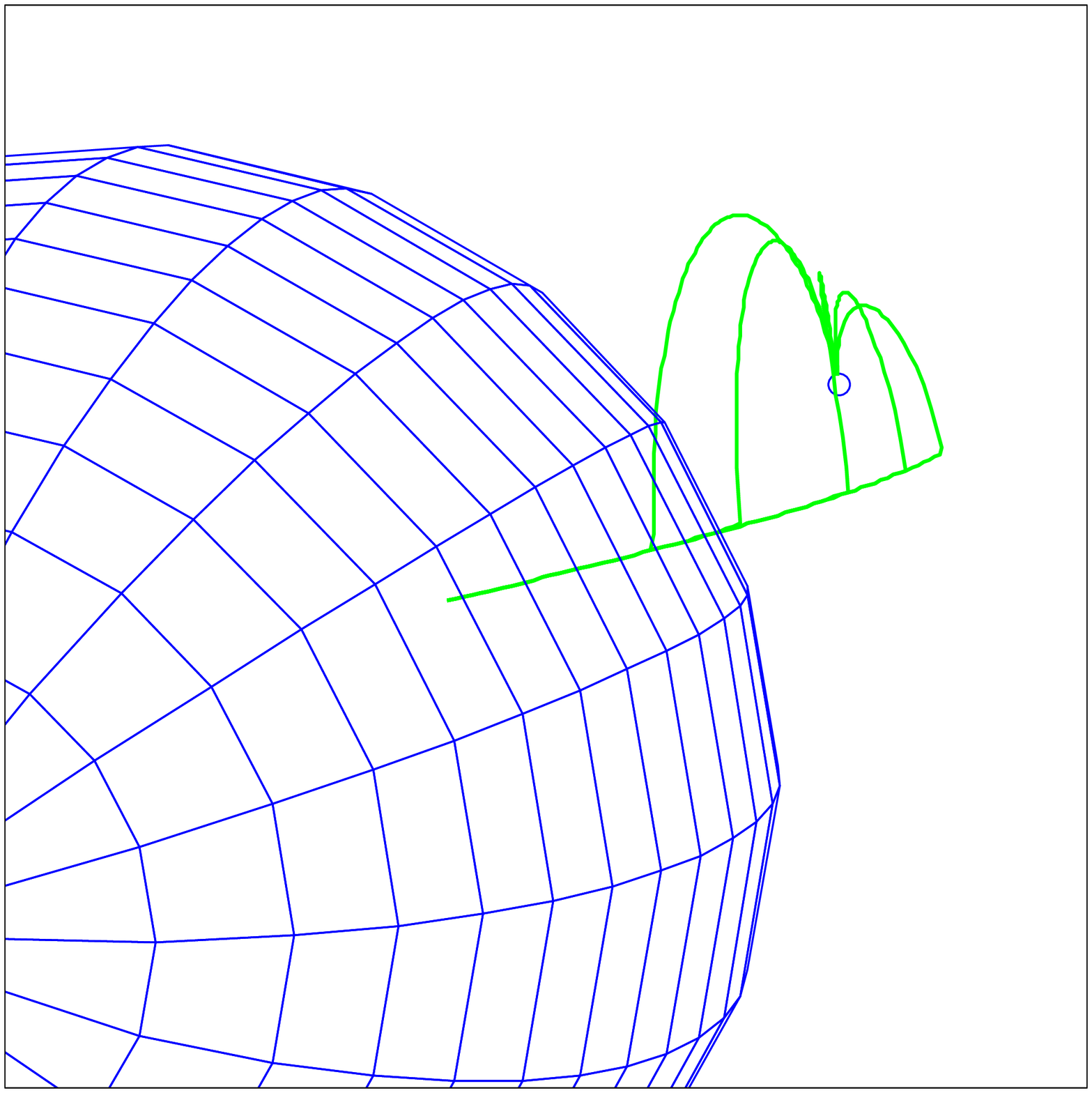,width=43mm,angle=0,clip=}
\end{minipage}\hfill
\begin{minipage}{43mm}
\psfig{figure=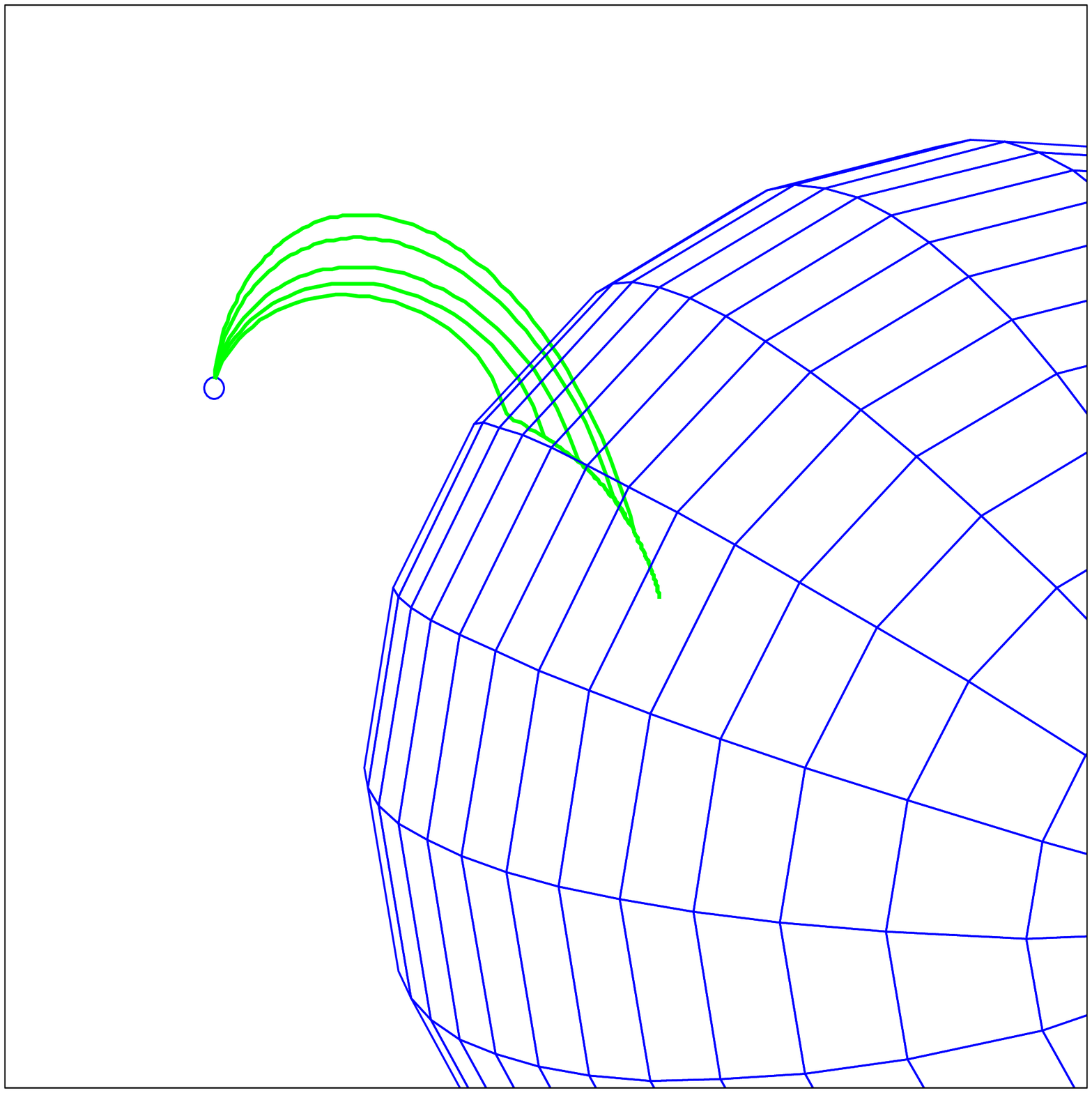,width=43mm,angle=0,clip=}
\end{minipage} 
\caption{The accretion geometry, which reproduces the main features seen in the Doppler tomograms and the light curves. In the upper left corner the view perpendicular to the orbital plane is shown, with the secondary star (right) and the accretion stream, falling onto the white dwarf (left). First it follows the ballistic trajectory, later it couples to the magnetic field and follows it to the magnetic pole. The upper right panel shows the same scenario in velocity coordinates, plotted over the tomogram of \heII. The lower panels show the system at phases $\phi=-0.06$ (left) and $\phi=0.068$ with the correct inclination angle.}
\label{f:mod_geo}
\end{figure}

With this model we interactively explored almost 
the whole parameter space within the limits on ($i, Q, K_1, K_2$) 
derived above. We assumed a dipolar magnetic field, it's orientation 
is described by the colatitude $\delta$ with respect to the rotation 
axis and the azimuth (or longitude) $\chi$ with respect to the line joining
both stars. The quadrant between $\chi = 0\degr$ and $90\degr$ 
contains the ballistic stream. 
A satisfactory qualitative fit to the spectroscopic and photometric features 
described above was reached for $\delta = 10\degr$, $\chi = -35\degr$, 
i.e.~with a nearly aligned dipole tilted  away from the ballistic stream.
Such a geometry in spatial and Doppler coordinates is reproduced 
in Fig.~\ref{f:mod_geo}.

Using these parameters we reach qualitative agreement between the observed
contact phases in broad-band and emission line light curves, of the
phasing of maxima in broad-band light curves, of the location of the
stagnation region at the terminal end of the ballistic stream in spatial
and Doppler coordinates and of the close vicinity of the magnetic and 
ballistic stream in Doppler coordinates.

The predicted location of the accretion spot, which is offset from the 
magnetic axis, is at stellar colatitude 17\degr~and at azimuth of $-16$\degr.
With this location of the spot we predict a maximum in the soft X-ray light 
curve (which is not yet observed) at binary phase $\phi = 0.045$. 
In the case of a pointlike accretion spot, a self-eclipse would be observed around phase $\phi = 0.55$.
If the accretion region is significantly extended on the surface of the white dwarf or in height, it will only be partially eclipsed.

\section{Discussion and conclusions}

The main results of our photometric and spectroscopic study with full 
phase coverage, paying particular attention to the eclipse properties 
is a re-determination of the system parameters of the binary. 
These are listed in Table~\ref{t:syspar}.

\begin{table}[htb]
\small
\caption{The system parameters of \v13, as derived in this work.
}{\label{t:syspar}}
\begin{center}
\begin{tabular}{|l|l|}
\hline
Parameter&Value\\
\hline
$P$ 			& $0.33261194(8)$ d\\
$\Delta t_{\rm ecl, wd}$& $2418\pm60$ s\\
$i$ 			& $76.6\degr - 78.9\degr$\\
$Q = M_1/M_2$ 		& $1.37 - 1.63$\\
$M_2$ 			& $0.6 - 0.4$ \msun\\
$M_{\rm wd}$ 		& $0.82 - 0.58$ \msun\\
$K_2$ 			& $180 - 200$ \kmps\\
$K_{\rm wd}$ 		& $110 - 145$ \kmps\\
$T_{wd}$ 		& $\le 20000$\,K\\
$D$ 			& $550 \pm 75$ pc\\
$\delta$		& $10$\degr\\
$\chi$ 			& $-35$\degr \\ 
\hline
\end{tabular}
\end{center}
\end{table}

The orbital period derived here from broad band photometry
has much higher accuracy 
than previously, since a larger time base could be used by us compared
to others. 
We found no evidence for an asynchronous rotation of the white dwarf.
The time difference between the centre of eclipse of the white 
dwarf and of the accretion stream could be determined for one epoch
by making use of an archival HST-observation to be 
172 ($\pm$ 20) s. 

A more accurate determination of the eclipse timings and of the eclipse 
duration of the accretion spot can probably obtained from upcoming 
XMM-observations. Based on our geometrical model we predict no 
self-eclipse of the accretion region. We found no evidence of a second
accretion region to be present but cannot certainly exclude its presence.

Based on the measured velocities of the irradiated front- and non-irradiated 
back-side of the secondary star, we could narrow the possible range of
masses of the secondary to $0.4-0.6$\,\msun. This is the first dynamic 
mass determination of the secondary in \v13\ and confirms or even lowers
previous mass estimates based on the spectral type. The secondary could
be resolved in the Doppler image of Na absorption lines, where 
it shows a marked depletion on the irradiated side. 

Doppler maps of the emission lines are clearly different to those
found in the literature for other eclipsing polars (HU Aqr, UZ For).
We could explain these differences with an extreme orientation of the 
magnetic axis pointing away from the ballistic stream. 

\begin{acknowledgements}
We thank B.~G\"{a}nsicke (G\"{ottingen}) for his grid of white
dwarf model atmospheric spectra. 
Furthermore, we thank our referee, P. Szkody, for useful comments which 
helped to improve the paper.

This work was supported by the DLR under grant 50 OR 9706 8 and the 
DFG under grant Schw 536/6--1.
\end{acknowledgements}

\end{document}